\documentclass[aps,prx,twocolumn,10pt,floatfix,superscriptaddress,showkeys,showpacs,twoside] {revtex4-2}

\usepackage{graphicx}

\usepackage{amsmath,amssymb,amsfonts}
\usepackage{mathtools}
\usepackage{bm}

\usepackage{ulem}
\usepackage{hyperref}

\newcommand{\ie}{{\textit{i.\,e.\@}}}
\newcommand{\eg}{{\textit{e.\,g.\@}}}
\newcommand{\cf}{{\textit{cf.\@}}}

\newcommand{\rmd}{\ensuremath{\mathrm{d}}}

\usepackage{xcolor}

\newcommand{\eq}[1]{(\ref{eq:#1})}
\newcommand{\Eq}[1]{Eq.~\eq{#1}}
\newcommand{\EQ}[1]{Equation~\eq{#1}}
\newcommand{\Eqs}[1]{Eqs.~\eq{#1}}

\newcommand{\fig}[1]{\ref{fig:#1}}
\newcommand{\Fig}[1]{Fig.~\fig{#1}}
\newcommand{\Figs}[1]{Figs.~\fig{#1}}
\newcommand{\FIG}[1]{Figure~\fig{#1}}

\newcommand{\tableRef}[1]{\ref{tab:#1}}

\newcommand{\TAB}[1]{Table~\tableRef{#1}}

\newcommand{\sect}[1]{\ref{sec:#1}}
\newcommand{\Sect}[1]{Sect.~\sect{#1}}

\newcommand{\app}[1]{\ref{app:#1}}
\newcommand{\App}[1]{App.~\app{#1}}

\newcommand{\za}{\ensuremath{\alpha}}
\newcommand{\zb}{\ensuremath{\beta}}
\newcommand{\zg}{\ensuremath{\nu}}

\newcommand{\moment}{\ensuremath{q}}
\newcommand{\qmom}{\ensuremath{q^{\text{th}}}} 

\begin{document}


\title{Universal hyper-scaling relations, power-law tails, and data analysis\\ for strong anomalous diffusion}

\author{J\"urgen Vollmer}
\thanks{ORCID: \href{https://orcid.org/0000-0002-8135-1544}{0000-0002-8135-1544}}
\affiliation{\text{Institute of Theoretical Physics,
    Universit\"at Leipzig, 
    Br\"uderstr.~16,
    D-04103 Leipzig, Germany}}

\author{Claudio Giberti}
\affiliation{\text{Department of Science and Engineering Methods, \emph{and}
    International Center INTER-MECH-Mo.Re,} 
    \text{Universit\`a di Modena e Reggio E., 
    Via Amendola 2, Padiglione Morselli, I-42100 Reggio E., Italy}}

\author{Jordan Orchard}
\affiliation{\text{Department of Mathematics, School of Science, Swinburne University of Technology, Victoria 3122, Australia}}

\author{Hannes Reinhard}
\affiliation{\text{Institute of Theoretical Physics,
    Universit\"at Leipzig, 
    Br\"uderstr.~16,
    D-04103 Leipzig, Germany}}

\author{Carlos Mej\'{i}a-Monasterio}
\thanks{ORCID: \href{https://orcid.org/0000-0002-6469-9020}{0000-0002-6469-9020}}
\affiliation{\text{School of Agricultural, Food and Biosystems Engineering, Technical University of Madrid,}
  \text{Av. Complutense s/n, 28040 Madrid, Spain}}

\author{Lamberto Rondoni}
\thanks{ORCID: \href{https://orcid.org/0000-0002-4223-6279}{0000-0002-4223-6279}}
\affiliation{\text{Department of Mathematical Sciences, 
    Politecnico di Torino, 
    Corso Duca degli Abruzzi 24, 
    I-10129 Torino, Italy}}
\affiliation{\text{INFN, Sezione di Torino, Via P. Giuria 1, 10125 Torino, Italy}}

\begin{abstract}
  Strong anomalous diffusion is {often} characterized by a piecewise-linear spectrum of the moments  of displacement.
  The spectrum is characterized by slopes $\xi$ and $\zeta$ for small and large moments, respectively,
  and by the critical moment $\za$ of the crossover.
  The exponents $\xi$ and $\zeta$ characterize the asymptotic scaling of the bulk and the tails of the probability distribution function of displacements, respectively.
  Here, we adopt asymptotic theory to match the behaviors at intermediate scales.
  The resulting constraint explains how distributions with algebraic tails imply strong anomalous diffusion,
  and it relates $\za$ to the corresponding power law.
  Our theory provides novel relations between exponents characterizing strong anomalous diffusion,
  and it yields explicit expressions for the leading-order corrections to the asymptotic power-law behavior of the moments of displacement.
  They provide the time scale that must be surpassed to clearly discriminate the leading-order power law from its sub-leading corrections.
  This insight allows us to point out sources of systematic errors in their numerical estimates.
  Rather than separately fitting an exponent for each moment we devise a robust scheme to determine $\xi$, $\zeta$ and $\za$.
  The findings are supported by numerical and analytical results on five different models exhibiting strong anomalous diffusion.
\end{abstract}


\maketitle


\section{Introduction}
\label{sec:intro}
In its simplest form, a transport process $x(t)$ is said to exhibit anomalous diffusion
if its Mean-Square Displacement (MSD) does not grow linearly in time. 
In recent years, great theoretical efforts have been made to unveil the physical mechanisms
that can induce anomalous diffusion (see, \eg~Refs.~\cite{Metzler,KRS08, Sok12} and references therein).
This has been accompanied by the rapid increase of highly-resolved experimental observations, particularly of video microscopy and particle tracking,
as well as by more refined numerical simulations of ensembles of trajectories \cite{frontiers}.
Processes exhibiting anomalous diffusion are observed in a broad range of phenomena including
molecules moving in the living cell \cite{Saxton,GC06},
dynamics on cell membranes \cite{NHB07},
crowded environments \cite{Sok12},
soil transport \cite{Rina},
heat transport in low-dimensional systems \cite{LLP}, and
in certain classes of billiards \cite{Ott, JBR08},
to mention a few.

\begin{figure*}
  \[
    \includegraphics{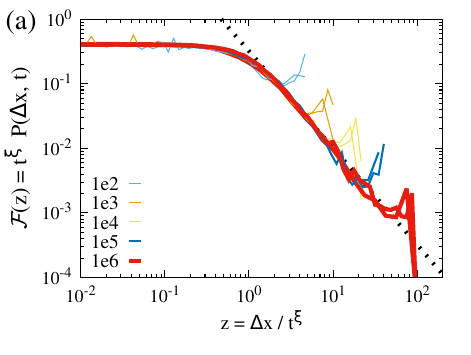}
    \qquad\qquad
    \includegraphics{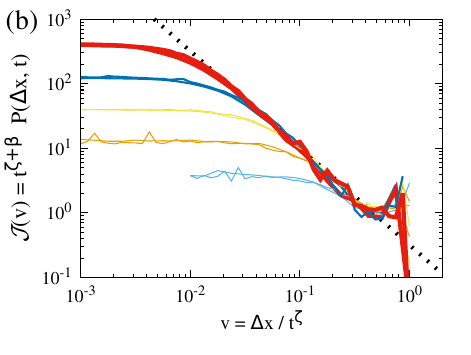}
  \]
  \caption{\label{fig:LLgPDFscaling}
    Scaling of (a) the bulk and (b) the tails of the PDF for the LLg with $\za=0.5$.
    Separate lines with the same color show the results of simulations with different ensembles. 
    They indicate the scatter of the numerical data.
    The color of the lines refers to times $t$, as provided in the legend of left panel.
    The dotted lines show an asymptotic power-law scaling for large~$z$ and small $v$, respectively.
    In the present case the power law has an exponent of $-(1+\za) = -3/2$,
    while $\xi=2/3$ and $\zeta = 1$.
    Details on the model and on numerical parameter vales are provided in \Sect{LLg}.
  }
\end{figure*}

In these  processes the probability distribution function (PDF) of the
displacement is  commonly non Gaussian, meaning  that the signatures
of anomalous diffusion appear beyond  the MSD, essentially in all the
moments  of the  distribution. Deviations  from the  normal Gaussian
behavior are  observed in the  tails and, in certain  dynamics, the
large fluctuations  of the  tails scale  differently from  the small
fluctuations described by the bulk of the distribution.

In  general, the time scaling  of the
$\qmom$ moment of displacement can be written as
\begin{subequations} \label{eq:spectrum}
{\begin{align} \label{eq:spectrum-evolution}
  \left\langle  \left\lvert x(t)-x(0) \right\rvert^\moment\right\rangle
  \sim  t^{\moment\, \zg_\moment /2} \ ,
\end{align}
} \hskip -3pt
$\zg_\moment = 1$ indicating normal diffusion as derived from Fick's law.
Conversely, anomalous diffusion corresponds to $\zg_\moment \ne 1$.
It comprises scale-invariant transport processes for which the anomalous behavior of the MSD extends to all moments of the displacement.
When $\zg_\moment$ varies with the moment order $q$, scale invariance breaks down,
and transport is said to be strongly anomalous \cite{CMMGA99}. 
In the literature, the most commonly investigated strong anomalous situation is characterized by two regimes, in which the power of $t$ is given by two different expressions, that are linear in the moment order:
\begin{align} \label{eq:momentPiecewise}
  \frac{\moment\, \zg_\moment}{2} = \left\{
  \begin{array}{lll}
    \xi \, \moment                               & \text{ for}  & \moment \leq \za \, ,
    \\[1mm]
    \zeta \, \moment - \za \, (\zeta - \xi )  & \text{ for}  & \moment > \za \, ,
  \end{array}
  \right .
\end{align}%
\end{subequations}%
where $\xi$, $\zeta$ {and $\za$} are given parameters.
Indeed, strong anomalous diffusion repeatedly arises as a combination of anomalous diffusion determining a local scale invariance for the small fluctuations of the displacement
(with $\moment \le \za$),
and rare but long-lasting ballistic excursions determining a faster scaling for the large fluctuations for which $\zeta=1$.
The piecewise linear behavior of \Eq{momentPiecewise} has been observed in the transport in polygonal billiards \cite{SL06,JR06,Orchard21},
billiards with infinite horizon \cite{AHO03,SS06,CESZ08},
diffusion in laser-cooled atoms \cite{AKB17}, 
one-dimensional maps~\cite{Pik91,SRGK15},  and
intermittent  maps~\cite{CMMGA99,AC03},
in experiments on the mobility of particles inside living cancer cells \cite{GW10}, 
the motion of particles passively advected by dynamical membranes \cite{SJ94},
running sand  piles~\cite{CLNZ99},
and in stochastic models  of inhomogeneous media~\cite{BCV10,BVLV14},
among a host of many other phenomena.
A great deal of insight into the strong anomalous diffusion of these processes has been gained
by means of generalizations of the central limit theorem in terms of non-normalizable densities
\cite{RDHB14a,RDHB14b,FSBZ15,ZDK15}.

The above scenario is not unique. For instance, 
in the bulk-mediated diffusion on lipid bilayers of Ref.~\cite{KCNP16},
the spectrum of moments is not piece-wise linear.
In the present paper, however, we investigate the double linear scaling of \Eq{momentPiecewise}, for a set of different deterministic and stochastic systems.
That is, in fact, the most commonly reported situation in experimental observations and theoretical studies.

Here we revisit the theory of strong anomalous diffusion
from the point of view of scaling theory \cite{Barenblatt} and asymptotic matching \cite{BO}
of the PDF $\mathcal{P}(\Delta x, t)$ for the displacements 
$\Delta x(t) = |x(t) - x(0)|$
over a time $t$.
Our main result is anticipated in \Fig{LLgPDFscaling}, which represents numerical data for the 
L\'evy-Lorentz gas (LLg), a simple but paradigmatic model of anomalous transport, which is described in 
\Sect{LLg}. 
Roughly speaking, \Fig{LLgPDFscaling} suggests that the process $\Delta x(t)$ evolves on two different asymptotic scales, 
which are revealed through the two scaled variables $z := \Delta x(t)/t^\xi$ and $v:=\Delta x(t)/t^\zeta$, concerning two different regions
of space.
More precisely, the idea numerically verified in \Fig{LLgPDFscaling} 
is the following.
For sufficiently large times $t$, the PDF is subdivided in two branches, 
corresponding to two distinct asymptotic regimes, which coincide over an interval that gets 
larger and larger as $t$ grows.
This is formalized through the following \emph{ansatz}:
\begin{align}\label{eq:ProbScalingCutoff}
  \mathcal{P}(\Delta x, t) \, \rmd \Delta x
  = \left\{
  \begin{array}{rll}
    \mathcal{F}(z) \, \rmd z
    & \text{with} 
    & z = \frac{\Delta x}{t^\xi} \leq c\, t^{\iota-\xi} \, ,
    \\[2mm]
    g(t) \, \mathcal{J}(v) \, \rmd v
    & \text{with} 
    & v = \frac{\Delta x}{t^\zeta} > c \, t^{\iota-\zeta} \, ,
    \end{array}
\right .
\end{align}
where $\xi,\iota,\zeta$ and $c$ are four parameters obeying
$0<\xi<\iota<\zeta$, 
$c >0$,
and the scaled variables $z$ and $v$ concern the ``bulk''  and the ``tail'' of $P$,
respectively defined by $\Delta x \in [0, c t^\iota)$ and $\Delta x \in (c t^\iota , \infty)$.
The functions $\mathcal{F}$ and $g \mathcal{J}$ are non-negative and integrable on their respective domains,
so that $P$ is non-negative and normalized over the half real line, for all~$t$.
The domain of $\mathcal{F}(z)$ grows without bounds with $t$,
since the value of $z$ at its right border, $c\,  t^{\iota-\xi}$, approaches $\infty$ in the large time limit.
Moreover, the domain of $\mathcal{J}(v)$ extends to arbitrarily small values,
since the smallest admissible value  $c\, t^{\iota-\zeta}$ for $v$ approaches $0$ for $t \to \infty$.

In \Fig{LLgPDFscaling} the scaling function $\mathcal{F}(z)$ takes the form of a power law for $z > z_c \simeq 5$,
and $\mathcal{J}(v)$ follows a power law for $v < v_c \simeq 0.2$.
We will show that as  a consequence 
$\mathcal{P}(\Delta x, t)$ follows a power law for
$ z_c \, t^\xi \lesssim \Delta x \lesssim  v_c \, t^\zeta $.
The two branches of $\mathcal{P}(\Delta x, t)$ overlap on this interval,
and the interval is expanding for increasing $t$:
The PDF features asymptotic matching
and, as long as $\xi < \iota < \zeta$ and $t$ is large,
the matching position for the two branches, $\Delta x = c  t^\iota$
will reside well inside this interval,
irrespective of the particular choice of $c$ and $\iota$. 

In \Sect{moments} we show how the piecewise linear scaling of $\zg_\moment$
emerges as a consequence of \Eq{ProbScalingCutoff}
and the asymptotic matching on the power law.
We establish universal relations between exponents characterizing the distribution,
and derive the leading-order corrections to the asymptotic power laws. 
These findings suggest a new framework for data analysis.
In \Sect{discussion} we explore these findings for five different models
that feature anomalous diffusion.
Besides elaborating on the LLg we revisit the exact solutions for L\'evy walks (LW) \cite{Metzler,ZDK15},
the fly-and-die (FnD)  \cite{VRTGM21}
and the slicer-map (SM) \cite{SRGK15} dynamics,
as well as numerical data on the Lorentz gas with 
infinite horizon (LGi) \cite{AHO03},
and polygonal billiard channels (PBC) \cite{SL06,JR06,SS06,Orchard21}.
Their distributions and spectrum of displacements are all described by the theory exemplified above through the LLg.
In \Sect{conclusion}, a summary is given of our main findings 
and an outlook to future perspectives.

\section{Scaling and Asymptotic Matching}
\label{sec:moments}

In this section we discuss in general the scaling of the moments of displacement $\Delta x$ at time $t$
determined by a PDF enjoying power-law behavior and asymptotic matching, 
as described above in the special case of the LLg.
In particular, we assume that at large times $t$ the time evolution of the bulk of the probability density obeys scaling with a scaling variable $z = \Delta x / t^\xi$
(\cf~\Eq{ProbScalingCutoff}),
and that it assumes a power law for large displacements,
\begin{subequations}\label{eq:match}
\begin{align}
\label{eq:ProbTailX}
  \mathcal{P}(\Delta x, t) \simeq \za \, k \, (\Delta x)^{-\za-1} \, t^{\za\,\xi} \, .
\end{align}
%

\subsection{Asymptotic Matching}
Given the definition \eq{ProbScalingCutoff}, the power law \eq{ProbTailX} may be written for the two branches of the PDF.
The bulk is 
described by the scaling function
\begin{eqnarray}
  \mathcal{F}(z)
  & \simeq & \za\, k \, (\Delta x)^{-\za-1} \, t^{\za\,\xi} \: \frac{\rmd \Delta x}{\rmd z}
  \nonumber \\ \label{eq:ProbTailZ}
  & = &
    \za\, k \, z^{-\za-1} 
    \qquad 
    \text{ for } z > z_c 
\end{eqnarray}
where the parameter $z_c$ is deduced from data.
For the tail, we rely on the function $g$ introduced in \Eq{ProbScalingCutoff},
noting that \Eq{ProbTailX} implies:
\begin{align} \label{eq:ProbV}
  \mathcal{J}(v)
  \simeq
  \frac{\za k (\Delta x)^{-\za-1}}{g(t)} \:  t^{\za \xi} \: \frac{\rmd \Delta x}{\rmd v}
  = \frac{\za k (\Delta x)^{-\za-1}}{g(t)}  \, t^{\za\,\xi + \zeta} \, .
\end{align}
Then, taking 
\begin{align} \label{eq:betadef}
  g(t) = t^{-\zb}  \quad \text{ with } ~~ \zb = \za \, (\zeta-\xi)
\end{align}
one obtains
\begin{align}
  \mathcal{J}(v)
  = \za\, k \, v^{-\za-1} 
\quad 
 \text{ for } v < v_c  
\label{eq:ProbTailV}
\end{align}
\end{subequations}
where $v_c$ is deduced from data, like $z_c$ is. 
At late times the bulk and the tail of the distributions overlap as power laws in a wide interval
whose width diverges asymptotically in time.

There are two important things to note here:

Firstly, in terms of the scaled variables and distribution,
the deviation from the power law can be
pushed below any desired bound by selecting larger values of $z_c$,
and smaller values for $v_c$.
Even when $v_c \ll z_c$ it suffices to wait for
large values of $t$  
until the interval
$z_c \,  t^\xi < \Delta x <  v_c \, t^\zeta$
is non-empty.
Eventually, in the large-$t$ limit, the length of the interval will diverge
because $\xi < \zeta$. 
This suffices for our investigation,
since this interval determines the asymptotic transport properties,
\ie~the scalings with time of the moments.

Secondly, we note that in \Eqs{ProbTailZ} and \eq{ProbTailV} not only the exponent $-(\za+1)$ of the power laws, 
but also the prefactors, $\za k$, agree.
This holds for by every set of exponents $\{ \za, \, \zb, \, \xi, \, \zeta \}$
characterizing the PDF,
and it will become the cornerstone of the data analysis presented in \Sect{parameterFitting}.

\subsection{Moments}

The $\moment^{\text{th}}$ moment of the displacement is defined by
\begin{align*}
\left\langle \bigl ( \Delta x \bigr )^\moment \right\rangle_t
    &:= \int_{0}^\infty \: \bigl( \Delta x \bigr)^\moment \: \mathcal{P}(\Delta x, t) \: \rmd \Delta x \, .
\end{align*}
For large $t$, where $z_c t^\xi < c t^\iota < v_c t^\zeta$, the integral can be broken into three pieces
\begin{align}
    \left\langle \bigl( \Delta x \bigr )^\moment \right\rangle_t
    =&  \; \int_{0}^{z_c t^\xi} \: \bigl( \Delta x \bigr)^\moment \: \mathcal{P}(\Delta x, t) \: \rmd \Delta x 
      \nonumber\\[5pt]
    &+ \; \int_{z_c t^\xi}^{v_c t^\zeta} \: \bigl( \Delta x \bigr)^\moment \: \mathcal{P}(\Delta x, t) \: \rmd \Delta x
      \nonumber\\[5pt]
   &+ \; \int_{v_c t^\zeta}^\infty \: \bigl( \Delta x \bigr)^\moment \: \mathcal{P}(\Delta x, t) \: \rmd \Delta x  \, .
\end{align}
In view of \Eq{ProbScalingCutoff}, in which the Jacobians of the coordinate transformations are already accounted for, and applying \Eq{ProbTailX},
the $\qmom$ moment can be approximated asymptotically in time as follows
\begin{align}\label{eq:momentSplitting}
    \left\langle \bigl ( \Delta x \bigr)^\moment \right\rangle_t
    \simeq &  \; t^{\moment\,\xi} \: \int_{0}^{z_c} z^\moment \: \mathcal{F}(z) \: \rmd z  
             + t^{\moment\,\zeta - \zb} \int_{v_c}^{\infty}  v^\moment \: \mathcal{J}(v) \: \rmd v
      \nonumber\\[6pt]
    &+ \za \, k \, t^{\za \xi} \: \int_{z_c\,t^\xi}^{v_c\,t^\zeta}  \: (\Delta x)^{\moment-\za-1} \: \rmd \Delta x \, .
\end{align}

The first two integrals take fixed finite values that do not depend on time. 
Moreover, for $\moment \neq \za$ the latter integral in \Eq{momentSplitting} can be expressed as a
difference of the two power laws with exponents 
that appear in the prefactors in front of the other two integrals, respectively.
Collecting terms with the same time dependence provides
\begin{subequations}
\begin{align} \label{eq:momentQ} 
  \left\langle \bigl ( \Delta x \bigr)^\moment \right\rangle_t
  = N_\moment \: t^{\moment\,\xi} + I_\moment \: t^{\moment\,\zeta-\zb}
    \quad\text{for}\quad \moment \neq \za  
\end{align}
with
\begin{align} 
  N_\moment
  &= \frac{\za \, k \, z_c^{q-\za}  }{\za - \moment}
  + \int_{0}^{z_c} z^\moment \: \mathcal{F}(z) \: \rmd z \, ,
  \\[2mm]
  I_\moment
  &= \frac{\za \, k \, v_c^{\moment-\za} }{\moment-\za}
  + \int_{v_c}^{\infty} v^\moment \: \mathcal{J}(v) \: \rmd v \, .
\end{align}
\end{subequations}

On the other hand, for $\moment = \za$ the third integral in \Eq{momentSplitting} takes the form of a logarithm,
\begin{align*} 
  \left\langle \bigl ( \Delta x \bigr)^\za \right\rangle_t
    =& \;\; t^{\za\,\xi} \: \int_{0}^{z_c} z^\za \: \mathcal{F}(z) \: \rmd z  
      + t^{\za\,\xi} \int_{v_c}^{\infty}  v^\za \: \mathcal{J}(v) \: \rmd v 
      \nonumber\\[6pt]
    &+ \za \, k \, t^{\za \xi} \; \ln\frac{v_c \, t^\zeta}{z_c \, t^\xi} \, .
\end{align*} 
This provides
\begin{subequations} \label{eq:momentAlpha}
\begin{align}  
  \frac{ \left\langle \bigl ( \Delta x \bigr)^\za \right\rangle_t }{ t^{\za\,\xi} }
  = N_\za + I_\za + \za\, k \: (\zeta-\xi) \: \ln t
\end{align}
where
\begin{align} 
  N_\za &= - \za \, k \: \ln z_c + \int_{0}^{z_c} z^\za \: \mathcal{F}(z) \: \rmd z \, ,
  \\
  I_\za &= \za \, k \: \ln v_c + \int_{v_c}^{\infty}  v^\za \: \mathcal{J}(v) \: \rmd v \, .
\end{align}
\end{subequations}

In either case the coefficients $N_\moment$ and $I_\moment$ take finite values as long as $\left\langle \bigl ( \Delta x \bigr)^\moment \right\rangle_t$ exists.
For finite $t$ this is always the case, if the distribution has a cut-off,
i.e.~when the speed of moving objects is bounded.
and the distribution has a cut-off.
Moreover, in the scaling regime these values do not depend on the choice of $z_c$ and $v_c$.
For instance, for all values $z_2 > z_1 \gg z_c$  
we have
\begin{align*} 
  N_\moment( z_2 ) &- N_\moment( z_1 )
  =
    \frac{\za \, k \, \Bigl( z_2^{q-\za} -z_1^{q-\za} \Bigr)   }{\za - \moment}
    + \int_{z_1}^{z_2} z^\moment \: \mathcal{F}(z) \: \rmd z
  \\[2mm]
  &= \frac{\za \, k \, \Bigl( z_2^{q-\za} -z_1^{q-\za} \Bigr)   }{\za - \moment}
    + \za \, k \; \int_{z_1}^{z_2} z^{\moment-\za-1} \: \rmd z
    \\
    &= 0
\end{align*}
and analogous arguments apply for $I_\moment$, $N_\za$, and $I_\za$.

It is useful to note the geometric interpretation of the coefficients $N_\moment$ and $I_\moment$.
To this ends we observe that normalization of the PDF implies that $\mathcal F$ is integrable for small~$z$.
A fortiori, this integrability at small $z$ applies to $z^q \mathcal F$ 
for all positive moments~$\moment$.
Moreover, for $\moment < \za$ the power-law tail is integrable for $z\to\infty$, and
$\za \, k \, z_c^{\moment-\za} = -(\moment - \za) \; \int_{z_c}^{\infty} \za\, k \, z^{\moment-\za-1} \: \rmd z$.
In this case, $N_\moment$ amounts to the integral of $z^\moment \: \mathcal{F}(z)$ over $\mathbb R_+$.
It takes a positive value.
In particular the normalization of the PDF entails the normalization of $\mathcal{F}(z)$ on $\mathbb R_+$,
\begin{align} \label{eq:Fnormal}
  1 = \int_{0}^ \infty \mathcal{F}(z) \, \rmd z \, .
\end{align}

On the other hand, for $\moment > \za$ the power law is integrable from zero to $z_c$, and
$\za \, k \, z_c^{\moment-\za} = (\moment-\za) \; \int_0^{z_c} \za\, k \, z^{\moment-\za-1} \: \rmd z$.
In this case $N_\moment$ amounts to the integral of $z^\moment \, \mathcal F(z) - \za\, k \: z^{q-\za-1}$.
The integral can also be taken over $\mathbb R_+$ also for  $\moment > \za$ 
when $\mathcal F(z)$ converges for instance 
exponentially to the power law for large values of $z$.
In any case $N_\moment$ takes a finite value as long as the $\moment$th moment exists,
and it can take positive and negative values.
 
Similarly, for $\moment > \za$ the coefficient  $I_\moment$ is positive and it amounts to the integral of $v^\moment \: \mathcal{J}(v)$ over $\mathbb R_+$,
while for $\moment < \za$ it amounts to the integral of the difference  $v^\moment \: \mathcal{J}(v) - \za\, k \: v^{q-\za-1}$.
In this case $\mathcal{J}(v)$ must converge sufficiently rapidly to the power law such that the integral can be performed starting from $v=0$,
and the sign of $I_\moment$ will then signify which portion of $\mathcal{J}(v)$ lies below or above the power law.
Similarly, the signs of $N_\za$ and $I_\za$ will also depend on the specific shape of $\mathcal F$ and $\mathcal J$, respectively. 

\begin{figure*}
  \[
    \includegraphics{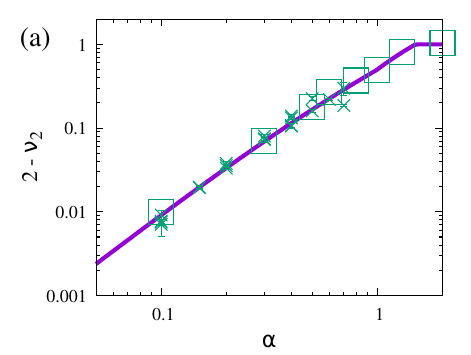}
    \qquad\qquad
    \includegraphics{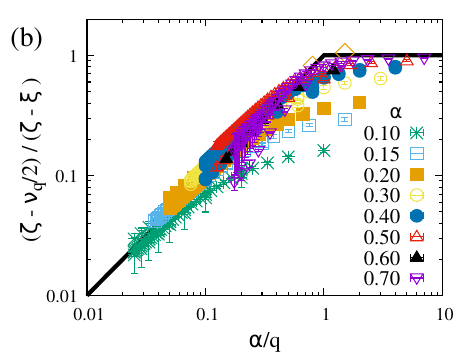}
  \]
  \caption{\label{fig:identifyParameters}
    (a) The parameter dependence of the MSD $\zg_2$ for the LLg,
    and (b) a plot striving for a data collapse of all moments on a universal scaling curve.
    The parameter values $\za$ for the right plot are specified in the legend of the panel.
    The solid lines show the analytical results of \Eqs{LLgMSD} and \eq{momentDataCollapse}, respectively.
    The symbols in the left panel were obtained from a fit to power law of the MSD (squares)
    and from the time scaling of the PDF of the displacement (crosses).
    Details on numerical parameters are provided in \Sect{LLg}.
  }
\end{figure*}

There are three scenarios for the asymptotic scaling of the moments:
$\moment < \za$, $\moment = \za$, and $\moment > \za$.

In the first case, 
the exponent $\moment \zeta - \zb$ is smaller than the exponent $\moment \xi$: 
\begin{align*}
  \moment \zeta - \zb = \moment \zeta - \za (\zeta - \xi)
  <
  \moment \zeta - \moment (\zeta - \xi) = \moment \xi \, .
 \end{align*}
Consequently, at the large $t$, the right hand side of \Eq{momentQ} is dominated by the addend containing $N_\moment$.
Conversely, in the third case, the exponent $q \zeta - \zb$ is larger than the exponent $q \xi$,
and
the large $t$ limit of the right hand side of \Eq{momentQ} is dominated by the addend containing $I_\moment$.
For $q=\za$ the term involving the logarithm dominate the right hand side of \Eq{momentAlpha}.
Asymptotically in time, we may thus write: 
\begin{align} \label{eq:momentsScaling}
  \left\langle \bigl ( \Delta x \bigr )^\moment \right\rangle_t
  &\simeq \left\{
      \begin{array}{lrl}
        N_\moment \; t^{\moment \, \xi}           & \quad\text{for}  &   0 < \moment < \za \, ,
        \\[2mm]
        \za \, k \, (\zeta-\xi) \: t^{\za \, \xi}  \; \ln t & \quad\text{for}  &  \moment = \za \, ,
        \\[2mm]
        I_\moment \; t^{\moment\, \zeta - \zb}     & \quad\text{for}  &   \za < \moment \, .
      \end{array}
      \right. 
\end{align}
This result is purely based on the constraints imposed by matching the power laws of the bulk and of the tail of the distribution.
In \Sect{discussion} we present numerical results to confirm that this theoretical framework 
applies to a vast variety of cases.

\subsection{Comparison with Numerical Data}
\label{sec:checkMoments}

In applications one is often interested in the scaling of the MSD, $\moment = 2$.
In \Fig{identifyParameters}(a) we show this data for the LLg where $\zeta=1$.
To this end we plot $2 - \zg_2$ as function of $\za$.
Normal diffusion, $\zg_2 = 1$, amounts to $2-\zg_2 = 1$,
and ballistic flights, $\zg_2 = 2$, amount to $2-\zg_2 = 0$.
In particular, for the LLg the MSD follows the exponent of normal diffusion when $\za>2$,
and the impact of ballistic flights become more and more noticeable when $\za$
decreases. 
It is then tempting to rewrite \Eq{momentPiecewise} in the form of a universal scaling curve
that admits a data collapse for all moments and all parameter values of a variety of models showing strong anomalous diffusion, with two linear regimes for the momenta spectrum:
\begin{align} \label{eq:momentDataCollapse}
  \frac{ \zeta - \frac{ \zg_\moment }{2} }{ \zeta - \xi }
  = \left\{
  \begin{array}{lrl}
    \za / \moment   & \quad\text{for}     & \za / \moment < 1 \, ,
    \\[2mm]
    1                  & \quad\text{for}     & \za / \moment \geq 1 \, .
  \end{array}
  \right .
\end{align}%
\FIG{identifyParameters}(b) shows this representation for a vast collection of moments
that we obtained by fitting LLg data (as discussed in \Sect{LLg} below).
The agreement is disappointing,
in spite of the fact that we have extremely clean numerical data (right panels of \Fig{LLgDistributions}). 
Close inspection reveals that the MSD appears to be a lucky borderline case.
For larger moments it becomes ever more complicated to faithfully sample the impact of the very long flights
that dominate the behavior of the large moments even though they are extremely rare.
For small moments $\moment$ there are severe systematic errors in the fits.

In order to understand the systematic errors in the fitted exponents
we explore the corrections to the asymptotic power laws.

\begin{figure*}
  \[
    \includegraphics[width=0.45\textwidth]{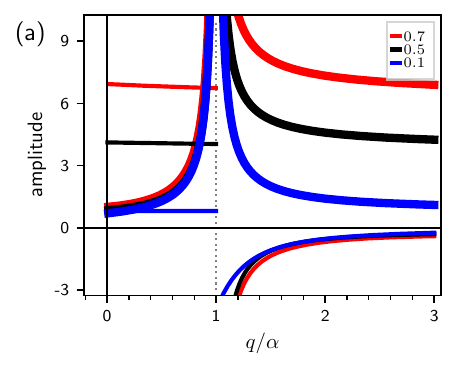}
    \qquad\qquad
    \includegraphics[width=0.45\textwidth]{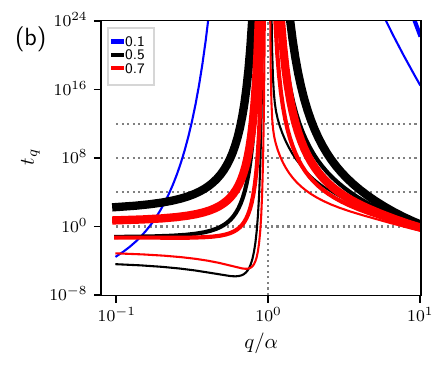}
  \]
  \caption{\label{fig:scaling-corrections}
    (a) The red, black, and blue lines show the prefactors of the power laws in \Eq{momentQ} for the LLg
    that have been calculated based on a fit of the asymptotic distributions $\mathcal F(z)$ and $\mathcal J(v)$.
    The thick lines refer to the leading order term, and the thin ones to the correction, respectively.
    Panel (b) provides the resulting estimates for the cross-over time $t_\moment$ as provided by \Eq{critTimeScale}.
    Thin, medium and thick lines refer to $\Phi = 1$, $3$, and $10$, respectively.
  }
\end{figure*}

\subsection{Corrections to Scaling}
\label{sec:fitProblems}

\EQ{momentAlpha} establishes
that there are logarithmic corrections to the power-law behavior for $\moment = \za$.
The resulting very weak time dependence features prominently in the discussion of the moments of displacement for the Lorentz gas with infinite horizon \cite{B92,AHO03,CGLS14b,CESZ08},
and in the discussion of the asymptotic scaling of its PDF \cite{D12,ZPFDB18}.
This will be addressed in \Sect{LG}.

To assess systematic errors in fitting exponents for $\moment \neq \za$,
we observe that \Eq{momentQ} predicts
that the $\moment$th moment of displacement will always be a superposition of two power laws.
The red, black, and blue lines in \Fig{scaling-corrections}(a) provide the prefactors of the power laws
for $\za=0.1$, $0.5$, and $0.7$, respectively.
The thick lines provide the prefactor of the dominant contribution, $N_\moment$ for $\moment<\za$ and $I_\moment$ for $\moment>\za$;
the thin lines the prefactor of the subdominant contribution.

\begin{figure*}
  \[
    \includegraphics{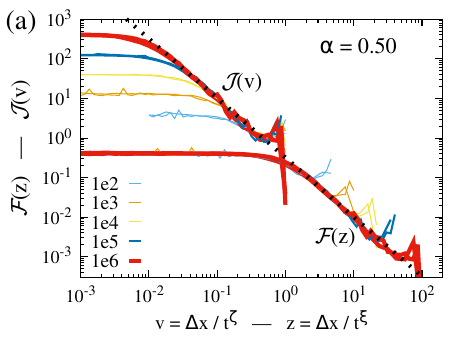}
    \qquad
    \includegraphics{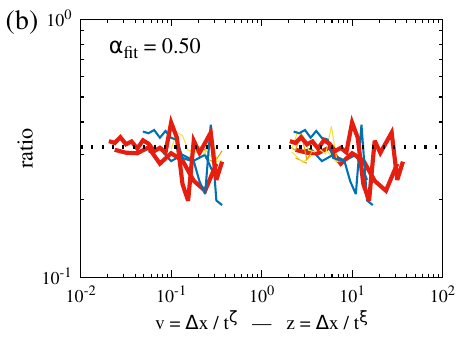}
  \]
  \[
    \includegraphics{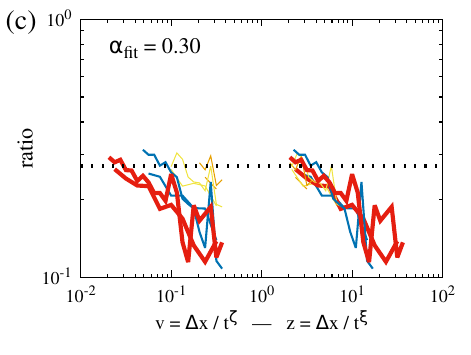}
    \qquad
    \includegraphics{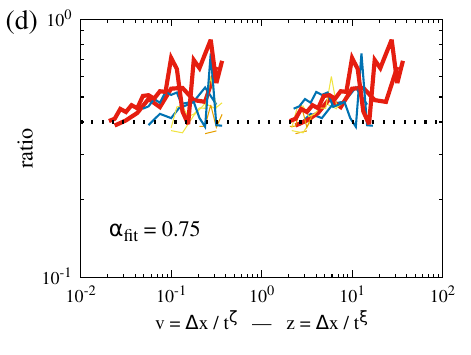}
  \]
  \caption{\label{fig:LLgTest}
    Simultaneous scaling of the bulk and the tails of the PDF of a LLg
    for the data also shown in \Fig{LLgPDFscaling},
    i.e.~for a parameter value where one expects $\za=1/2$, $\xi=2/3$, $\zeta=1$,
    and $\zb = \za \: (\zeta-\xi) = 1/6$. 
    (a)~Joint plot of the data of both panels \Fig{LLgPDFscaling}.
    (b)~A reduced plot showing the ratio of the data and the pertinent values of the power law that is given by the dotted line in panel~(a).
    The panels~(c) and~(d) show reduced plots of the best fits when one rather adopts the values $\za=0.3$ and $\za=0.75$, respectively.
    Note that adopting a different value of $\za$ alters the power of $t$ in the definition of $\mathcal J(v)$,
    and the slope of the dotted line in panel~(a).
    Details on the model and on numerical parameter values are provided in \Sect{LLg}.
  }
\end{figure*}

The scaling functions $\mathcal F$ of the LLg proceed entirely below the power law (\cf~\Figs{LLgPDFscaling}, \fig{LLgTest}(a) and \fig{LLgDistributions}).
Hence, $N_\moment$ takes a negative value for $\moment > \za$,
and the correction takes the form of a power law
that decays towards the asymptotic scaling from below.
Consequently, the fitted exponents are systematically too large
as long as the subdominant power law in \Eq{momentQ} is not substantially smaller than the leading order.
This is in line with the observation reported in \Fig{identifyParameters}(b)
that the fitted exponents are systematically too large
(\ie~the data points tend to lie below the solid black line).

The leading-order correction to the asymptotic power-law scaling involves the ratio of the two prefactors,
\begin{subequations}\label{eq:momentScaling}%
  \begin{align}%
    \frac{ \left\langle \bigl ( \Delta x \bigr )^\moment \right\rangle }{ I_\moment \, t^{\moment\zeta-\zb} }
    & \simeq 
      1 + \left( \frac{N_\moment}{I_\moment} \right) \; t^{(\za - \moment) (\zeta - \xi)}
      \quad\text{ for } \quad \za < q \, ,
    \\[5pt]
    \frac{ \left\langle \bigl ( \Delta x \bigr )^\moment \right\rangle }{ N_\moment \, t^{\moment\xi} }
    & \simeq
      1 + \left( \frac{I_\moment}{N_\moment} \right) \;  t^{(\moment - \za) (\zeta - \xi)}
      \quad\text{ for } \quad \za > q \, .
  \end{align}%
\end{subequations}%
For a reliable estimate of the exponents we demand
that the leading order term is larger than the correction term by a factor $\Phi$.
This requires
that the power-law fits are performed for times well beyond the critical time $t_\moment$ defined by:
\begin{subequations}
\begin{align} \label{eq:critTimeScale}
  & t_\moment = \left( \Phi \: N_\moment / I_\moment \right)^\tau
\quad \text{for}\quad   \za < q
  \\[5pt]
   & t_\moment = \left( \Phi \: I_\moment / N_\moment  \right\rvert^\tau 
 \quad \text{for}\quad   \za > q
\end{align}
with
\begin{align*} 
  \tau
  = \frac{1}{ \lvert \moment - \za \rvert \: (\zeta-\xi) } \, .  
\end{align*}
\end{subequations}%
\Fig{scaling-corrections}(b) provides the estimates for the LLg.
Thin, medium and thick lines refer to $\Phi = 1$, $3$, and $10$, respectively,
and the colors indicate different values of $\za$.
For $\za = 0.1$ and $\Phi=1$ (thin blue line) the values of $t_\moment$ are larger than $10^{24}$ for a noticeable range of the plot,
for $\Phi=3$ the time $t_\moment$ drops below $10^{24}$ only for $\moment \gtrsim 10\,\za = 1$,
and for $\Phi=10$ it remains substantially larger than that value for all considered values of $\moment$.
On the other hand, the largest times that we reach in the simulations is $10^9$.
Consequently, the systematic error in the fits is large,
and this is reflected in the severe deviations between prediction and numerical estimates
that we saw in \Fig{identifyParameters}(b).
For $\za = 0.5$ the maximum time in our simulations was $10^6$.
In this case the numerical estimates are expected to be reliable for $\za/\moment \lesssim 1/3$,
and this is indeed what we observe in \Fig{identifyParameters}(b).
For the $\za = 0.7$ the situation should even be better.
However, in that case the numerical estimates suffer from bad statistics,
which is reflected in large error bars. 

The time scale $t_\moment$ is model specific, but the discussion of the LLg reveals that it can easily be very large.
In particular, estimates of moments of the displacement will be hard
when the difference between $\xi$ and $\zeta$ is small,
or when $\moment$ is close to the cross-over moment $\za$.

In order to avoid these issues we propose to determining the exponents $\zeta$, $\xi$, and $\za$ based on a scaling analysis of the PDF of the displacement $\Delta x$.

\subsection{Data Analysis of the PDFs} 
\label{sec:parameterFitting}

In order to extract the scaling exponents directly from the scaling of the PDF we 
note that for small arguments $\mathcal{J}(v)$ takes exactly the same functional form as $\mathcal{F}(z)$ for large arguments
(cf.~\Eqs{ProbTailZ} and \eq{ProbTailV}).
In \Fig{LLgTest}(a) this is demonstrated by providing the scaling plot of the bulk and the tails of the PDF
in the same coordinate system, rather than plotting them separately as in \Fig{LLgPDFscaling}.
One clearly sees now that the large $z$ data for $\mathcal{F}(z) = t^\xi \mathcal{P}(\Delta x,\,t)$ lie on the same power law
as the small $v$ data of $ \mathcal{J}(v) = t^{\zeta+\zb} \mathcal{P}(\Delta x, \, t)$.

The three exponents characterizing the distribution can therefore be found by a data collapse of the
time evolution of the PDFs of the displacement:
\begin{itemize}
\item[$\xi$]  is used to collapse the height of the bulk distribution $\mathcal{F}(z) = t^\xi \, \mathcal{P}(\Delta x, t)$ for small~$z = \Delta x/t^\xi$.
  This one parameter fit will provide a data collapse for the bulk and the power-law tail of the distribution.
  It is robust because it involves a vertical shift of horizontal lines such that they all meet at the same level.
  For the data in \Fig{LLgTest} this provides the estimate $\xi = 0.66\pm 0.02$ for a theoretically expected value of $\xi=2/3$.
  The resulting data collapse of the data is also shown in 
  \Fig{LLgPDFscaling}(a).
  
\item[$\zeta$] is used to collapse the position of the cutoff value~$v_{\text{max}}$
  for the tails of the distribution $\mathcal{J}(v) = t^{\zeta+\zb} \, \mathcal{P}(\Delta x, t)$
  for the largest admissible values of~$v$.
  The vertical position of the data is adjusted by the choice of $\zb$.
  The fit of $\zeta$ is robust because it involves the horizontal shift of very sharp peaks such that they all lie at the same height;
  or rather the minima at the end of the algebraic decay should lie at the same height.
  Moreover, in many circumstances $\zeta$ need not be fitted because it must be unity
  when the large-scale cutoff is dominated by ballistic flights.
  One obtains a plot of the data as shown in 
  \Fig{LLgPDFscaling}(b).
  
\item[$\za$] must provide the slope to the \emph{simultaneous} asymptotic scaling of \emph{both}
  the large $z$ asymptotics of $\mathcal{F}(z)$ and
  the small $v$ asymptotics of~$\mathcal{J}(v)$.
  According to \Eq{betadef} it is related to $\zb$ by the hyperscaling relation $\zb=\za\,(\zeta-\xi)$.
  A joint plot of both distributions is shown in \Fig{LLgTest}(a).
  The quality of the fit can best be judged via a reduced plot where one plots the deviation from the suggested power law.
  The power-law scaling for $\mathcal{F}$ applies for $z \gtrsim z_c = 5$
  and the scaling for $\mathcal{J}$ applies for $v \lesssim v_c = 0.2$.
  A reduced plot of the data in this range is shown in \Fig{LLgTest}(b).
  The exponent $\zb$ must be adjusted such that all data points have the same value.
  The effect of varying $\za$ for fixed $\zeta$ and $\xi$ is demonstrated in \Fig{LLgTest}(c) and (d). 
\end{itemize}

One clearly sees that the \emph{simultaneous} matching of the slopes of \emph{both} line segments provides significant better grips on the data than a mere power-law fit.
In the next section we show that this data analysis provides a reliable characterization of numerical data.

\begin{table}
  \begin{align*}
  \begin{array}{l@{\quad}l@{\qquad}l@{\qquad}c@{\qquad}l}
    \hline
    \text{model}  & \text{parameter} & \xi            & \zeta  & \za      \\
    \hline
    \text{LLg}    & 0 < p_1 < 1      & (1+\za)^{-1}    & 1      & {p_1}        \\
                  & 1 < p_1          & 1/2            & 1      & 2 p_1 - 1    \\
    \text{FnD}, \text{SM}    &  0 < p_2      & 0      & 1      & p_2          \\
    \text{LW}     & 1 < p_3 < 2      & p_3^{-1}        & 1      & p_3          \\
    \text{LGi}    & 0 < p_4 < 1      & 1/2            & 1      & 2            \\
    \text{LGf}    & 1 < p_4          & 1/2            & 1      & \infty        \\
    \text{PBC}    & p_5 \approx 1.0649 & {\approx 0.5}  & 1    & \approx 2   \\
                  & p_5 \approx 0.935  & {\approx 0.5}  & 1    & \approx 3   \\
    \text{SDCA}   & 1 < p_6 < 2      & p_6^{-1}        & 3/2    & p_6 \\ 
    \hline
  \end{array}
  \end{align*}
  \caption{\label{tab:Spectrum}
    Choices of $\xi$, $\za$, and $\zb$ to match the spectrum of exponents for different model systems
    that characterize the anomalous scaling of the distribution, \Eq{ProbScalingCutoff},
    the algebraic tails of this distribution, \Eq{ProbTailZ} and \eq{ProbTailV},
    and the time-scaling of the cutoff function, \Eq{betadef}, respectively.
    Irrespective of the choice in the literature we denote the
    parameters used to define the dynamics as $p_1 , \cdots, p_6$ {(see
    section~\ref{sec:discussion})}.
  }
\end{table}

\section{Discussion of Specific Models}
\label{sec:discussion}

We discuss five prominent model systems featuring anomalous transport:
\begin{description}
\item[LLg] the L\'evy-Lorentz gas (LLg) \cite{BCV10,SRGK15}
\item[SM]  the slicer map (SM) \cite{SRGK15},
\item[LW]  L\'evy walks (LW) \cite{Metzler,ZDK15},
\item[LGi]  the Lorentz gas with infinite horizon \cite{AHO03}.
\item[PBC] polygonal billiard channels (PBC) \cite{SL06,JR06,SS06,Orchard21}.
\end{description}
The considered models feature vastly different values of the exponents $\xi$ and $\za$.
The exponents provided in the quoted literature for the respective models
are summarized in \TAB{Spectrum}. 
In addition we include in the table the values for the Lorentz gas with a finite horizon (LGf) \cite{D14},
and spatial diffusion of cold atoms (SDCA) \cite{AKB17}.
In the LGf the tails of the distribution decay faster than any power law ($\za = \infty$)
such that it can not feature strong anomalous transport (\cf~\App{fastCutoff}).
SDCA features power-law tails and a cutoff of the distribution
that scales with a non-common exponent $\zeta = 3/2$.
Still it shows all relations between the exponents that are established in the present study.

\begin{figure*}[p!]
  \[
    \includegraphics{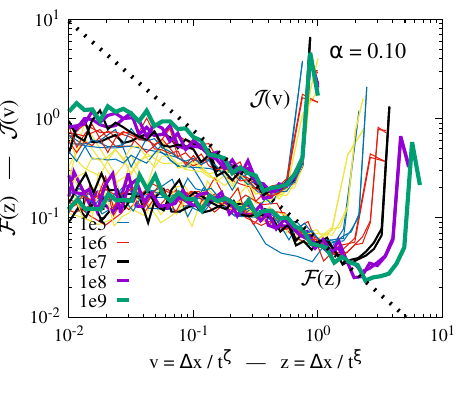}
    \qquad\quad
    \includegraphics{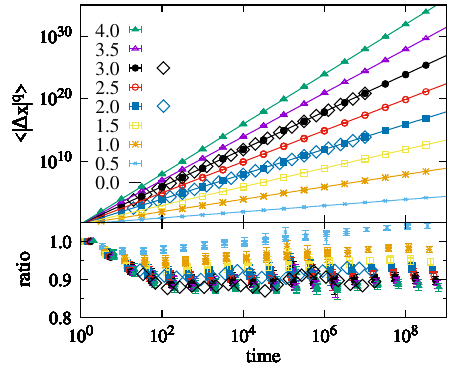}
  \]
  \[
    \includegraphics{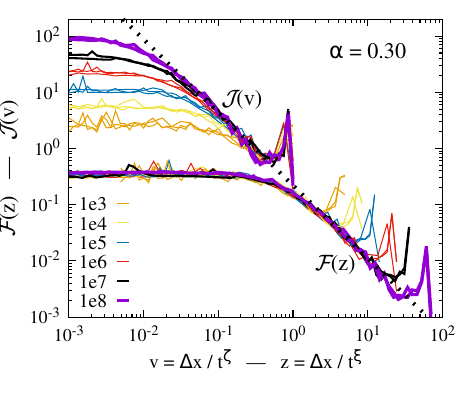}
    \qquad\quad
    \includegraphics{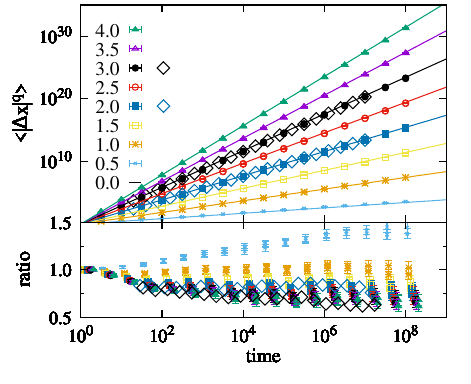}
  \]
  \[
    \includegraphics{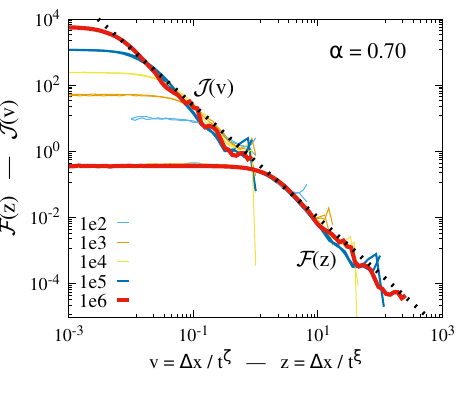}
    \qquad\quad
    \includegraphics{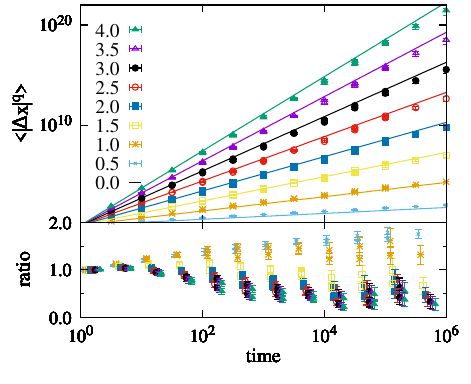}
  \]
  \caption{\label{fig:LLgDistributions}
    Compilation of numerical data for the LLg for parameters
    $p_1=\za=0.1$ (top), 
    $p_1=\za=0.3$ (middle), and
    $p_1=\za=0.7$ (bottom).
    The left panels show the joint data collapse for the bulk and the tail of the distribution
    where the power-law decay is indicated by a dotted line, as introduced and explained in \Fig{LLgTest}.
    The respective upper panels of the right show the power-law growth of different moments of the distribution,
    with color coding as indicated in the panels.
    Below each of these panels we provide the reduced plots for the moments,
    where one can clearly see the corrections to scaling,
    \Eq{momentScaling}.
  }
\end{figure*}

\subsection{The L\'evy-Lorentz Gas (LLg)}
\label{sec:LLg}

The LLg describes a one-dimensional ballistic motion of a particle
that moves with unit velocity and is scattered at point scatterers located at fixed, random positions in space.
The distance $r$ between neighboring scatterers is distributed according to a L\'evy-type distribution,
$f(r) \propto p_1 \, r^{-p_1-1}$.
The particle is transmitted and reflected with the probability $1/2$ when it encounters a scatterer.
The PDF $\mathcal{P}(\Delta x, t)$ for the displacement
$\Delta x$ for an ensemble of particles starting at $t=0$ on a given scatterer features strong anomalous diffusion.
Due to the fixed unit speed of the particles the probability $\mathcal{P}(\Delta x, t)$ vanishes for $\Delta x > t$.
The tail of the distribution has a cutoff that scales with $\zeta=1$.
The bulk of the distribution can be rescaled by a different exponent $\xi$,
and it has a power-law tail. 
For $p_1 < 1$ the bulk is rescaled by the exponent $\xi = 1/(1+\za)$ and its power law decays with exponent $\za = p_1$.
This non-trivial dependence of $\xi$ with parameter $p_1$ arises from a delicate interplay of spatial disorder and random scattering events,
as derived in \cite{BCV10}. 
For $p_1>1$ one finds $\za=2 p_1 -1$ and diffusive scaling $\xi = 1/2$ for the bulk distribution \citep{BCV10,SRGK15}. 
For the LLg, \Eq{momentPiecewise} entails
\begin{align}\label{eq:LLgMSD}
  2 - \zg_2 = \left\{
  \begin{array}{lrl}
    \displaystyle
    \za^2 / ( 1 + \za )
    &      \text{for} &  0 < \za < 1 \, ,
    \\  
    \ \za / 2 \;
    & \quad\text{for} &  1 \leq \za < 2  \, ,
    \\  
    \ 1 
    & \quad\text{for} &  2 \leq \za  \, .
  \end{array}
  \right .
\end{align}%
This is shown as the solid curve in 
\Fig{identifyParameters}(a).
The square symbols were computed from a fit to power law of the mean square
displacement as a function of time as used in \cite{GRTV17,SRGK15}.
Instead, the cross symbols were obtained from a proper scaling of the PDF of the displacement.
There is excellent agreement between the data and the theoretical prediction.

The scatter of the data in 
\Fig{identifyParameters}(b)
is due to the correction to the power laws reported in
\Eq{momentScaling}.  For the LLg the relations $\zeta = 1$ and
$\xi = 1/(1+\za)$ imply
\begin{align}
  \tau
  = \frac{1}{ |\moment - \za| (\zeta - \xi) }
  = \frac{1}{\za} \;\; \frac{1+\za}{ |\moment - \za |}  \, . 
\end{align}
Hence, the correction term decays slowly when $\moment \approx \za$ and for small $\za$.
This leads to the huge values of the cross-over time for $\za = 0.1$
that we observed in \Fig{scaling-corrections}(b),
and it accounts for the large values at $\moment \approx \za$ for $\za = 0.5$ and $0.7$.
Consequently, the fitted exponents feature a systematic bias towards larger exponents $\moment \zg_\moment /2$,
as clearly shown by the data in \Fig{identifyParameters}(b)
and the right panels in \Fig{LLgDistributions}.

The different panels of  \Fig{LLgDistributions}  show numerical data for parameter values
$p_1=\za=0.1$, 
$p_1=\za=0.3$, and
$p_1=\za=0.7$, respectively.
For $p_1=0.3$ the data analysis is relatively straightforward,
just as in the case  $p_1=\za=0.5$ discussed in \Fig{LLgTest}.
The other two values exemplify generic problems for large and small values of $\za$.

For slowly-decaying power-law tails, $\za=0.1$, the long ballistic flights give rise to substantial fluctuations in the bulk of the distribution.
One needs here a lot of statistics to perform stable averages to estimate $\xi$.
At the same time one needs very long simulations to resolve a noticeable range of the power-law tail.
Even for times as large as $10^9$ the power-law piece of the distribution only extends over half a decade.
Performing a simultaneous fit based on the scaling of $\mathcal{F}(z)$ and $\mathcal{J}(v)$ extends this range to two orders of magnitude,
such that a reliable estimate comes into reach.

For rapidly decaying power-law tails,  $\za=0.7$, it is a challenge to collect data with sufficient statistics
to faithfully resolve the power-law tails.
However, for $\mathcal{F}(z)$ and $\mathcal{J}(v)$ the data with a good statistics lie in vastly different regions in the mutual plot;
in the present example at $5 \lesssim z \lesssim 400$ and $6 \times 10^{-3} \lesssim v \lesssim 0.2$ for the $10^6$ data.

\begin{figure*}[p!]
  \[
    \includegraphics{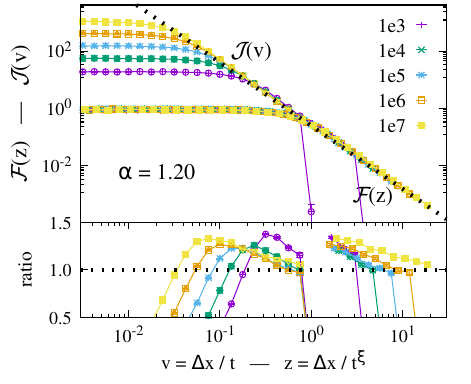}
    \qquad
    \includegraphics{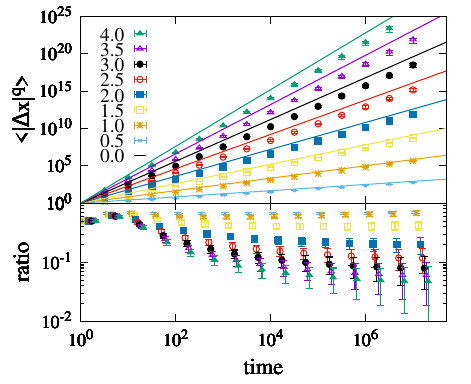}
  \]
  \[
    \includegraphics{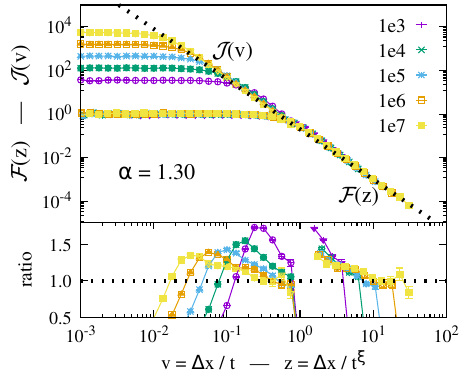}
    \qquad
    \includegraphics{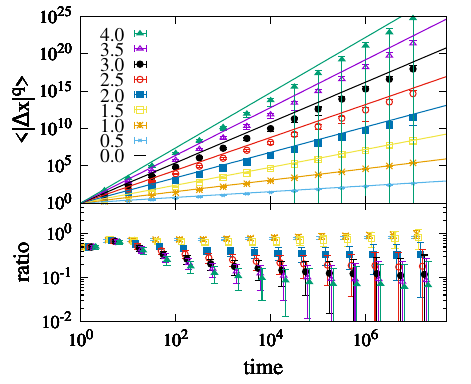}
  \]
  \[
    \includegraphics{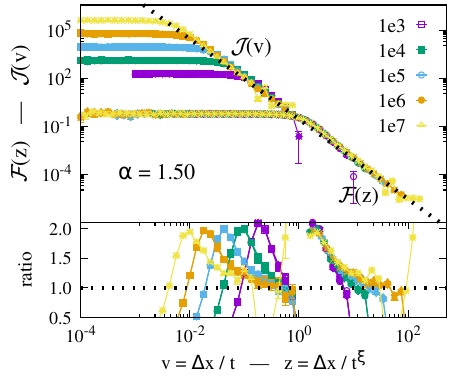}
    \qquad
    \includegraphics{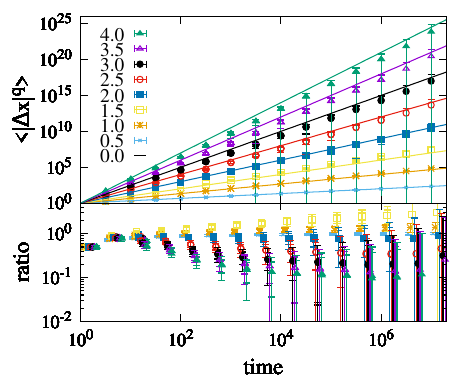}
  \]
  \caption{\label{fig:LFDistributions}
    Compilation of numerical data for the L\'evy walks for parameters
    $p_3=\za=1.2$ (top), 
    $p_3=\za=1.3$ (middle), and
    $p_3=\za=1.5$ (bottom).
    The left panels show the joint data collapse for the bulk and the tail of the distribution
    where the power-law decay is indicated by a dotted line, as introduced and explained in \Fig{LLgTest}.
    The respective upper panels of the right show the ratio of this data and the power-law fit of the tails of the distribution.
    Below each of these panels we provide the reduced plots for the moments,
    where one can clearly see the corrections to scaling, \Eq{momentScaling}.
  }
\end{figure*}

\subsection{Fly-and-Die (FnD) and Slicer Map (SM) dynamics}

The slicer map was introduced in \citet{SRGK15} as an analytically solvable dynamical system
that features all regimes of anomalous transport,
from localization to ballistic motion. In addition,
it was shown that the slicer map does also exhibit strong anomalous transport.
Its displacement time autocorrelation function was analytically obtained \citep{GRTV17},
and more recently, a universal scaling of the moments of the displacement
and of their correlation function was suggested  in Ref.~\citep{VRTGM21}.
The latter paper also introduces the FnD dynamics as a dynamical system where it is even more straightforward to perform the analytical calculations.
From a coarse-grained perspective, that focuses on the scaling of the distribution of the displacement,
both dynamics are characterized as follows.
The initial distribution is an approximation of a delta function at the origin,
and transport proceeds in a one-dimensional space where the displacement from the origin is denoted as $\Delta x$.
Each member  $x$ of the ensemble is characterized by a life time, $t_c(x)$.
For $t<t_c(x)$ it propagates with unit velocity.
Then it stops and remains at its terminal position $\Delta x = t_c(x)$ for all times $t>t_c(x)$,
with negligible fluctuations in the case of the SM.
The distribution of life times $t_c$ obeys a power law $\mathop{\text{Prob}}( t_c > t ) = (t/t_0)^{-p_2}$
where $t_0$ is a smallest admissible life time.

For $t > t_0$ the dynamics provide the PDF of displacements.
It can be written in the form of \Eq{ProbScalingCutoff}
with 
\begin{align*}
  \mathcal{F}(z) &= \left\{
  \begin{array}{llll}
    0
    & \text{ \ for  }
    & z = \frac{\Delta x}{t^0} < t_0 \, ,
    \\[1mm]
    \displaystyle
    \frac{p_2}{t_0^{-p_2}} \: z^{-p_2-1}
    & \text{ \ for \ }
    & t_0 <\!\!  z = \frac{\Delta x}{t^0} < t \, ,
  \end{array}
  \right .
  \\[1mm]
  g(t)\mathcal{J}(v) &= \left\{
  \begin{array}{llll}
    \displaystyle
    t^{-p_2}\: \frac{p_2}{t_0^{-p_2}} \: v^{-p_2-1} 
    & \text{ \ for} \quad \frac{t_0}{t^1} <  v = \frac{\Delta x}{t^1} < 1 \, ,
    \\[2mm]
    \displaystyle
    t^{-p_2} \: \frac{p_2}{t_0^{-p_2}} \: \delta( v - 1 ) 
    & \text{ \ for} \quad   v = \frac{\Delta x}{t^1} = 1 \, ,
    \\[1mm]
    0
    & \text{ \ otherwise \ }
   \end{array} 
  \right.
\end{align*}
such that $\xi=0$, $\zeta=1$, and $\za = \zb = p_2$.
In \citet{GRTV17}, the authors used the MSD to identify the relations between the parameters of the SM and the LLg,
and pointed out that this entails a matching of all other moments.
From the perspective of our present results this is an immediate consequence of~\Eq{momentsScaling}.

\subsection{L\'evy walks (LW)}

As a next step we verify that the findings established for the LLg also apply for L\'evy walks.
To this end we consider one-dimensional L\'evy walks
where particles move with unit velocity along a line.
Their  motion is similar to the one of the LLg.
However, in the LLg the particles move in a domain of scatterers that provide an environment with fixed (quenched) disorder,
while the distance of each segment of the LW is sampled randomly and independently \citep{1989BlumenZumofenKlafter,RDHB14a,ZDK15,DBSB15}.
We consider an ensemble where all particles start in the origin, 
and they move in a given direction for a time $t$
sampled from a L\'evy distribution $f(t) \propto p_3 \: t^{-p_3-1}$ with $t\in [1, \infty)$.
At the end of each step the particles reverse direction with probability $1/2$.

Arguably, LWs are the most elaborately studied model system featuring strong anomalous transport.
In particular, the different scaling of the bulk and the tails of the PDF have first been observed in this context,
and in those studies the distribution $J(v)$ in \Eq{ProbV} has been denoted as non-normalizable density
\citep{RDHB14a,RDHB14b,ZDK15}.
The moments of displacement have been reviewed in \cite{Metzler,ZDK15}.
They follow the behavior anticipated in \Eqs{momentPiecewise} with $\za = p_3$, 
$\zeta = 1$, and $\xi = 1/\za$.

The panels in  \Fig{LFDistributions} show numerical data for $p_3 = \za = 1.2$,  $1.3$, and $1.5$.

The simulations were run till time $10^7$,
with ensemble comprising more than $3.1 \times 10^5$, $9.2 \times 10^4$, and $1.2 \times 10^5$ trajectories for $\za = 1.2$,  $1.3$, and $1.5$, respectively.
The left panels show the PDF and the matching of the two branches of the intermediate power law.
Data points with different colors refer to data evaluated at different times, as indicated in the legend.
The upper right panels show data for the moments ranging from $0.5$ till $4.0$, in steps of $0.5$.
The lower ones show the ratio of the data and the analytical prediction for their asymptotic power law scaling.
These reduced plots reveal that even for this high-quality data it is challenging to determine the asymptotic scaling of the moments of the distribution by fitting the power laws.
On the other hand the matching of the intermediate power laws of the bulk and the tail of the distribution (lower left panels) provides robust results.

\subsection{The Lorentz gas (LG)}
\label{sec:LG}

The Lorentz gas is another classical model that features strong anomalous transport~\cite{B92,AHO03,D14}.
This model comprises the elastic scattering of a point particle in a two-dimensional setting
where circular scatterers of radius $R=1$ are arranged on a hexagonal or square lattice.
One deals with the displacement of an ensemble of initial conditions that start close to the origin,
and all move with unit speed.

The transport properties are studied as a function of the ratio $p_4$ of the radius $R$ of the scatterers
and the critical radius $R_c$ 
where channels that admit infinitely long collisionless flights are closed.

\begin{figure*}
  \[
    \includegraphics[width=0.43\textwidth]{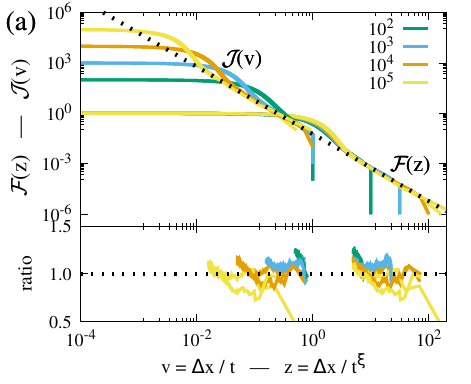}
    \qquad\quad
    \includegraphics[width=0.43\textwidth]{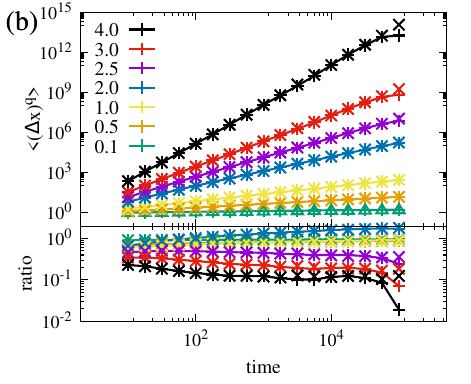}
  \]
  \caption{\label{fig:LG}
    Time evolution of the PDF of the displacement (a) and the moments of displacement (b) 
    for a Lorentz gas where the scatterers are arranged on a square lattice with parameter $p_4=0.3$.
    Different colors in (a) mark the simulation time $10^2, \cdots, 10^5$, as indicated in the legend of the plot,
    while the lines of corresponding color show the displacement along the two lattice direction of the arrangement of scatterers.
    In (b) the data for different lattice directions are marked by crosses and plusses,
    and different colors indicate the moment $\moment$ of $\left\langle \bigl ( \Delta x \bigr )^\moment \right\rangle_t$. 
  }
\end{figure*}

For $p_4 > 1$ the LG is said to have finite horizon (LGf). 
The case of a finite horizon emerges for Lorentz gases with a hexagonal arrangement of scatterers.
In such a setting the lengths of the flights between collision have a finite expectation value,
and the central limit theorem asserts that $\xi = 1/2$.
Moreover, when the mean-free path is bounded the PDF $\mathcal{P}(\Delta x, t)$ of the displacement also
does not have heavy tails.
As a consequence, a LG with finite horizon features normal transport \cite{D14}. 
In \App{fastCutoff} we show that within our theory such a system can not have power-law tails.

For $p_4 <1 $ infinitely long collisionless segments of trajectories can pass through the channels of the periodic arrangement of the scatterers,
the mean free path is no longer bounded and the length of the flights between collision do not have a finite expectation value \cite{B92}.
The billiard is said to have infinite horizon (LGi).
This gives rise to strong anomalous transport \cite{AHO03,CESZ08,CGLS14b}
where the spectrum \Eq{momentPiecewise} is obtained
for the choice $\za = 2$, $\zeta = 1$, and $\xi = 1/2$.
The value of $\zeta$ reflects the motion of the particles with unit speed,
and the value of $\za$ is in accordance with the finding of \cite{ZPFDB18}
that the PDF of displacements features a power-law tail that decays with an exponent $-3$.

The coincidence of the cross-over dimension $\za$ with the variance 
gives rise to mathematical challenges with the central limit theorem
and logarithmic terms in the asymptotic scaling of the distribution function \cite{MT16,ZPFDB18}.
Indeed, according to \Eq{momentAlpha} the variance must involve the observed $t\,\ln t$ scaling
when $\za = 2$.

In order to explore this dynamics we consider an arrangement of scatterers on a square lattice.
In this case $R_c$ amounts to half the lattice spacing.
\Fig{LG} shows numerical data for the case $p_4=0.3$.
The scaling of the bulk of the distribution is diffusive, $\xi=1/2$.
As expected for a system where the tails are due to ballistic trajectories that evolve with a fixed finite speed,
the cutoff for large displacements scales with $\zeta=1$,
Moreover, the crossover region between the bulk and the cutoff features a power-law tail with exponent $\za=2$.

\begin{figure*}
    \[
      \includegraphics[width=0.45\textwidth]{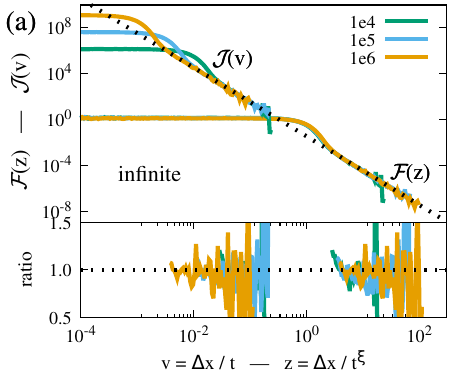}
      \qquad
      \includegraphics[width=0.45\textwidth]{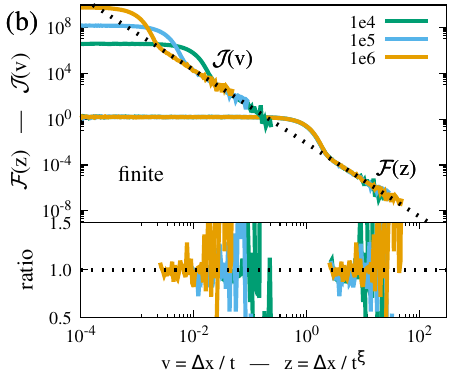}
    \]
    \caption{
      Scaling of the bulk and tails of the displacement distribution for polygonal billiard channels
      with the parameters $p_5 \approx 1.0649$ (a, infinite horizon) and $p_5 = 0.935$ (b, finite horizon).
      Details on the model and its geometric parameters are provided in \Sect{PBC}.
      For both panels, the scaling parameters $\xi = 1/2$ and $\zeta = 1$ are used.
      The dotted lines in the panels have the slopes
      (a) $-\za - 1 = -3.0$ and  
      (b) $- \za - 1 = -3.2$ for the infinite horizon and finite horizon, respectively.
      The lower panels show the ratio of the data and 
      power laws
      that are provided by the black dotted lines in the main panels.
      The moments of the distribution are discussed in \Fig{BPC_crossovers}.
    }
    \label{fig:PBC-Scaling}
\end{figure*}

The reduced plot of the power-law tails of the PDF (lower panel in \Fig{LG}(a)) supports the scaling of the tails with $\za=2$.
For the present data this exponent can only be found, however, by a fit based on the $\mathcal{F}(z)$ and the $\mathcal{J}(v)$ data.
The range of the power law in itself is too narrow for a reliable fit,
such that other approaches will provide a larger value for the exponent.
Moreover, the reduced plots for the time evolution of the moments (lower panel of \Fig{LG}(b)) are also not entirely horizontal.
For the entire range till times $10^5$ the reduced plot for the moment $q=0.1$ has a positive slope,
and for times $t<10^3$ the fourth moment has a negative slope.
The latter curve becomes reasonably horizontal for $t \simeq 10^4$ and then statistical errors become noticeable.
An accurate fit of the moments of this data will have to take into account the higher order correction terms, \Eq{momentScaling},
and they might have to account for logarithmic contributions to the scaling of the bulk of the distribution \cite{D12,ZPFDB18}.
Certainly the approach provides important additional insight on top of a brute forward analysis \cite{CGLS14b,CESZ08} of the moments of displacement.

\subsection{Polygonal billiard channels  (PBC)}
\label{sec:PBC}

In order to underline this point we explore polygonal billiard channels  (PBC),
\ie~a class of model systems where there is no theoretical insight concerning the values of the parameters characterizing their strong anomalous transport. 
A PBC consists of freely flying point particles
that experience specular reflections at two polygonal boundaries which form an infinitely long, narrow channel.
In recent years, a number of studies have investigated the transport properties of polygonal billiard channels, mostly via numerical simulations of the statistics of displacement \cite{JBR08, SL06, JR06, Orchard21, SS06, VRTGM21, ARV02}.
When the sides of the polygons are parallel
transport is anomalous with a MSD scaling faster than linear.
If the boundaries are not parallel, the MSD scales linearly in time as in normal diffusion.
In contrast Ref.~\cite{SL06} reported long transients of subdiffusive transport for walls which are close to parallel.
For PBC with finite and infinite horizon the transport is strongly anomalous,
with a ballistic scaling for moments of the displacement higher than a certain threshold order \cite{SL06, Orchard21,VRTGM21}.
When the horizon is infinite, 
it is expected that large fluctuations will be determined by long ballistic excursions.
However, for a finite horizon, polygonal channels also exhibit a ballistic scaling of the large fluctuations,
due to geometrically induced correlations and the lack of chaos \cite{Orchard21, VRTGM21}.
Moreover, transport in polygonal billiard channels was found to belong in some cases to the universality class of FnD dynamics \cite{VRTGM21},
with the biscaling of the PDF of displacement being reported in Ref.~\cite{VRTGM21}.
Further investigation on the statistics of first-passage time for channels of finite length showed a remarkable dependence on whether the polygonal angle is a rational or irrational multiple of $\pi$
(in the former case the distribution of the first-passage time is scale-free) \cite{Orchard21}.

Here we adopt the notation used in Ref.~\cite{JR06}, and revisit geometries previously studied in Ref.~\cite{SL06,VRTGM21}.
We considered two different channels with $\Delta x = 1, \Delta y_b = 0.45, \Delta y_t = 0.77$ and $H = 1.27$ yielding infinite horizon, and $H = 1.17$ yielding finite horizon.
Similarly to the Lorentz gas, we define a parameter $p_5 = (H - \Delta y_b)/ \Delta y_t$,
so that $p_5 < 1$ stands for a channel with finite horizon
and $p_5 > 1$ for a channel with infinite horizon.
For the two geometries considered here, $p_5 \approx 1.0649$ and $p_5 \approx 0.935$ respectively.

\begin{figure*}
  \[
    \includegraphics[height=0.35\textwidth]{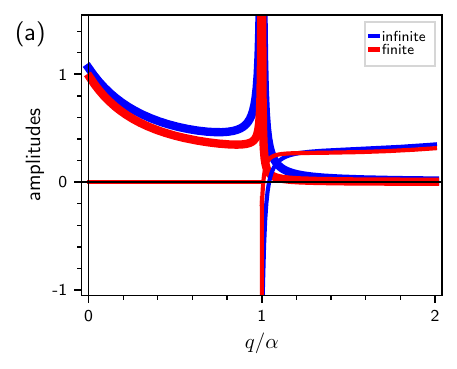}
    \qquad\quad
    \includegraphics[height=0.35\textwidth]{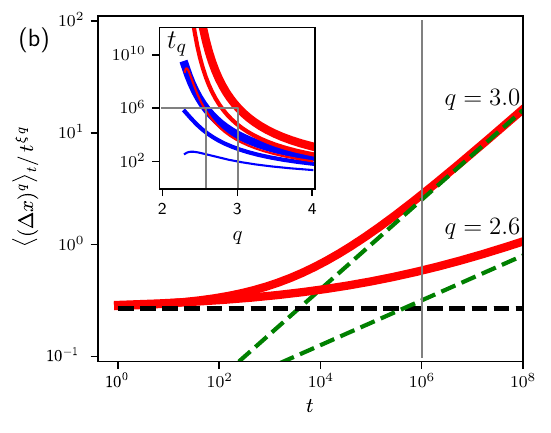}
    \]
    \caption{\label{fig:BPC_crossovers}
      Analogously to \Fig{scaling-corrections} panel (a) shows the prefactors of the two power laws governing the time evolution of the moments of displacement of the PBC. 
      In this case the sub-leading amplitudes are very small for $\moment < \za$,
      and they take positive values for values of $\moment$ slightly larger than $\za$.
      (b) The resulting the cross-over from the scaling of the moments with exponent $\xi\,q$ at early times (horizontal black dashed line) to the asymptotic behavior (green dashed lines) for the BPC with finite horizon and two different values for $\moment$.
    Simulations that run till time $10^6$ do not resolve the asymptotic scaling for $\za < \moment \simeq 3$.
    The red lines in the inset show the estimates for the cross-over time, $t_q$ (\cf~\Eq{critTimeScale}) for
    $\Phi = 1$ (thin lines), $\Phi = 3$, and $\Phi = 10$ (thick lines), respectively.
    The blue lines show the corresponding curves for the PBC with infinite horizon
    where the cross-over emerges much earlier.
  }
\end{figure*}

We follow the approach outlined in  \Sect{parameterFitting}  to determine the exponents governing the transport:
\begin{itemize}
\item
  $\xi = 1/2$  provides a collapse in the bulk of the distribution for both $p_5 < 1$ and $p_5 > 1$,
  as shown by the collapse of $\mathcal F(z)$ in the main panels of \Fig{PBC-Scaling}.
\item
  $\zeta = 1$ yields a collapse of the tail of the distribution.
  After all, particle move with finite speed
  such that the large fluctuations of the displacement obey a ballistic scaling, irrespective of $p_5$.
  This is shown by the collapse of $\mathcal I(v)$ shown in the main panels of \Fig{PBC-Scaling}.
\item
  The dotted lines in the main panels, and the reduced plots shown in the insets suggest 
  to select the values $\za \approx 2.0$ and $\za \approx 2.2$
  for the infinite horizon ($p_5 \approx 1.0649$, left panel)
  and the finite horizon ($p_5 \approx 0.935$, right panel), respectively.
\end{itemize}

The authors of Refs.~\cite{VRTGM21, SL06} analyzed numerical data based on power law fits of the moments of displacement.
For the billiard with an infinite horizon they report a value of $\za \approx 2$.
This is in line with the present findings.
In contrast, for the finite-horizon case they find a value of $\za \approx 3$,
that lies far outside the confidence interval of the data analysis performed in \Fig{PBC-Scaling}(b).
We believe that this is due to the systematic errors arising in the fitting the power laws
discussed in \Sect{fitProblems}. 
To underpin this point we estimated $N_\moment$ and $I_\moment$ based on our numerical data.
The inset of \Fig{BPC_crossovers}(b) shows the resulting dependence of the cross-over times $t_\moment$.
The main panel of the figure shows the cross-over for the finite-horizon PBC.
The vertical gray bar in that plot shows the maximum time of the simulations.
For $t \lesssim 10^6$ one only starts to see the asymptotic scaling when $\za \gtrsim 3$.
This implies that the data fitting for $\za < \moment \lesssim 3$ is performed in the sub-leading regime
that is still following the $\moment\xi$ scaling.
In contrast to the LLg the prefactor of the power law $N_{\moment}$ takes now positive values also for $\moment$ slightly larger than $\za$. 
As a consequence, a power-law fit to the moment of displacement will show exponents
that follow the small $\moment$ branch of the data.
The cross-over to the steeper branch is observed only for values of $\moment$ that lie substantially above $\za$.
Consequently, estimates of $\za$ based on fits of the moments of displacement tend to provide values that are systematically too large.
For the PBC with finite horizon this effect is particularly severe
because it features a very small value of $k$ in \Eq{match}. 
As a consequence, $N_\moment$ takes positive values outside a tiny neighborhood of $\za$,
and $I_\moment$ takes very small values for $\moment/\za > 1$.
This leads to a the cross-over time $t_\moment$ with exceptionally large values,
and numerical data on the moments of displacement are stuck in the sub-dominant scaling regime.

\section{Conclusion}
\label{sec:conclusion}
In this communication we employed methods of asymptotic matching
that have widely been used in the context of hydrodynamic boundary layers and other scaling theories.
Here, they have been used to deal with the large-distance cutoff of the probability density describing the PDF of displacement in strong anomalous diffusion.
We established that strong anomalous transport emerges whenever the PDF features algebraic tails.
Indeed in \App{fastCutoff} we provide the formal proof
that strong anomalous diffusion only arises if this distribution has power-law or more slowly decaying tails.

PDFs giving rise to strong anomalous diffusion are characterized by three exponents:
$\xi$ describes how the width of the distribution becomes broader in time;
$\zeta$ describes the scaling of the cutoff for large times;
$\za+1$ is the exponent of the power-law tails.
In the present paper we have fully characterized the scaling of the moments of the displacements 
via hyper-scaling relations that are based only on the moment $\moment$ and these exponents:
\begin{itemize}
\item
In \Eq{momentQ} we relate the time dependence of the moments of displacement to $q$, $\za$, $\zeta$, and~$\xi$.
This prediction is purely based on the constraints arising from matching the bulk and the tail of the distribution.
\item
  By \Eqs{ProbScalingCutoff} and
  \eqref{eq:momentPiecewise}, the slope $\xi$ of the lower moments of the distribution is related to the scaling of the bulk of the distribution.
  In most applications such a scaling is entailed by large-deviation theory.

\item
 By \Eqs{ProbScalingCutoff} and \eqref{eq:momentPiecewise} the slope $\zeta$ of the higher moments of the distribution is related to the scaling of the cutoff 
 of the distribution, which will always arise for processes with bounded speed.
  
\item
  By \Eq{ProbTailZ} the cross-over between the two regimes is connected to the exponent $\za$ characterizing the power-law tail of the PDF.
In \App{fastCutoff} we proved that all moments are dominated by the bulk scaling when the tail decays faster than a power law.
There is no strong anomalous diffusion in that case.

\item
  In \Eq{momentAlpha} we establish that the moments $\moment = \za$ grow like $t^{\za\xi} \: \ln t$.
  The relevance of this result for the moments of displacement of the Lorentz gas have been discussed in \Sect{LG}. 

\item
  There is also an exponent $\zb$ that describes the time scaling of the distribution 
  characterizing the tails.
  By \Eq{betadef} we express this exponent in terms of the other exponents $\za$, $\zeta$, and~$\xi$.

\item
  In \Eq{momentScaling} we provide the leading-order corrections to the power-law growth of the moment of the displacement.
  In \Eq{critTimeScale} we provide the time scale that must be surpassed to clearly discriminate the leading-order power law from its sub-leading corrections.
\end{itemize}

In \Sect{fitProblems} and by \Fig{identifyParameters}(b) and the right panels of \Figs{LLgDistributions}--\fig{LFDistributions} we demonstrate that slowly-decaying corrections to the power-law growth of the exponents will often give rise to severe systematic errors of estimated exponents.
An informed way to determine $\za$, $\zeta$ and $\xi$ is described in \Sect{parameterFitting}.
The impact of adopting this approach is demonstrated in \Sect{PBC}:
fitting of the moments provides an estimates of $\za$ that is too large by almost $40$\%
because it is virtually impossible to observe the asymptotic scaling for moments in the vicinity of $\za$
(see~\Fig{BPC_crossovers}(b, inset)).

In \Sect{discussion} the hyper-scaling relations have been verified for a range of vastly different systems.
To this end we revisited analytical results and and we presented some new numerical work on systems where no analytical results were available so far.
We conclude that our approach provides a novel unified perspective on analyzing the moments of displacement of strong anomalous diffusion.
It is based on hyper-scaling relations and a thorough analysis of the time scaling of the PDF for the displacement.

\begin{acknowledgments}
  %
JV thanks the \textsf{Politecnico di Torino} for an invitation as a distinguished visiting professor. 
This work was initiated during this visit.
JV and LR acknowledge very insightful discussions with  Konstantin Schwark, and Muhammad Tayyab.
JO acknowledges financial support from the Australian Government Research Training Program Scholarship.
C. M.-M. acknowledges financial support through the project grant PID2021-127795NB-I00 by MCIN/ AEI / 10.13039/501100011033 / FEDER, UE.
This research was performed under the auspices of Italian National Group of Mathematical Physics (GNFM) of INdAM.
\end{acknowledgments}

\section*{Author contributions}

JV came up with the idea, and discussed it with LR, CMM and CG.
He generated the LLg and LW data to test the idea and
set up the data analysis.
JO and HR contributed the PBC and LG data, respectively.
JV provided a first draft of the paper, and did the final editing.
CG, CMM, and LR contributed to preparing the manuscript.

\appendix

\section{Moments of displacement for weak anomalous diffusion}
\label{app:fastCutoff}
In this appendix we show that weak anomalous diffusion emerges
when the distribution $\mathcal P(\Delta x, t)$ 
features scaling for small values of $\Delta x/t^{\xi}$ and it
decays faster than any power for large $\Delta x$.
Scaling of the bulk of the distribution implies
that there are positive real numbers $\iota > \xi > 0$ and $c > 0$
such that for large times we have
\begin{subequations}
\begin{align}\label{eq:ProbScalingWeak}
  \mathcal{P}(\Delta x, t) \, \rmd \Delta x
  = \mathcal{F}(z) \, \rmd z
    \qquad \text{for} 
    \quad z = \frac{\Delta x}{t^\xi} < c\, t^{\iota-\xi} \, .
\end{align}
At the same time a decay of the tail that is faster than any power implies
that for every $\moment > 0$ there is a constant $k_\moment$ and a time $t_\moment$
such that 
\begin{align} 
  \mathcal{P}(\Delta x, t) < k_\moment \: t^{\xi} \; \lvert \Delta x \rvert^{-\moment-2}
  \qquad \text{ for } \quad t > t_\moment \, .
\end{align}
\end{subequations}

We consider the following lower bound, $L_\moment(t)$, 
for the $\moment$th moment of the distribution,
\begin{align*} 
  L_\moment(t)
  &=  \int_0^{c\, t^\iota} \rmd \Delta x \; |\Delta x|^\moment \: \mathcal{P}(\Delta x, t)
  \\[2mm]
  &\leq \Bigl\langle \left\lvert\Delta x\right\rvert^\moment \Bigr\rangle_t
    = \int_0^{\infty} \rmd \Delta x \; |\Delta x|^\moment \: \mathcal{P}(\Delta x, t)
  \\[2mm]
  &= \int_0^{c\, t^\iota} \rmd \Delta x \; |\Delta x|^\moment \: \mathcal{P}(\Delta x, t)
    + \int_{c\, t^\iota}^\infty \rmd \Delta x \; |\Delta x|^\moment \: \mathcal{P}(\Delta x, t)
  \\[2mm]
  &= L_\moment(t) + D_\moment(t) \, .
\end{align*}
It applies because the integrand is always positive. 

In the following we consider values $\iota > \xi$. 
In that case the time evolution of $L_\moment$ is governed by the scaling of the bulk of the PDF,
\begin{align*} 
  L_\moment(t) &= t^{\moment\,\xi} \: \int_0^{c\, t^{\iota-\xi}} \rmd z \; z^\moment \: \mathcal{F}(z)
  \to_{t \to \infty}
  N_\moment \: t^{\moment\,\xi} \, ,
\end{align*}
where 
$ 
  N_\moment = \int_0^\infty \rmd z \; z^\moment \: \mathcal{F}(z)
$ 
is a finite positive number if the moment exists.

Further we observe that the fast decay of the probability implies an upper bound on $D_\moment(t)$,
\begin{align*} 
  D_\moment(t)
  &< k_\moment \: t^{\xi} \; \int_{c\, t^{\iota}}^\infty \rmd \Delta x \; |\Delta x|^\moment \; |\Delta x|^{-\moment-2}
  = \frac{ k_\moment }{c} \: t^{\xi-\iota} 
\end{align*}
For large times this bound tends to zero:
the upper and lower bound converge in the long-time limit.
Hence, the moment 
must scale in time in the same way as $L_\moment(t)$, 
\begin{align*} 
  \Bigl\langle \left\lvert\Delta x\right\rvert^\moment \Bigr\rangle_t
  &= N_\moment \: t^{\moment\,\xi} \qquad \text{ for } \quad t \gg t_\moment \, .
\end{align*}
For $\xi = 1/2$ the system features normal diffusion, 
and otherwise there is weak anomalous diffusion.


%

\end{document}